\documentclass[preprint,showkeys,aps,prb]{revtex4}
\usepackage{amsmath}
\usepackage[dvips]{graphicx}
\usepackage{setspace}
\begin{document}

\title{
Ergodic Isoenergetic Molecular Dynamics
for Microcanonical-Ensemble Averages
}

\author{
William Graham Hoover and Carol Griswold Hoover       \\
Ruby Valley Research Institute  \\
HC 60 Box 601    \\
Ruby Valley, NV 89833                             \\
}

\date{\today}


\vspace{0.1cm}

\begin{abstract}
Considerable research has led to ergodic isothermal dynamics which can replicate
Gibbs' canonical distribution for simple ( small ) dynamical problems.  Adding one
or two thermostat forces to the Hamiltonian motion equations can give an ergodic
isothermal dynamics to a harmonic oscillator, to a quartic oscillator, and even to the
``Mexican-Hat'' ( double-well ) potential problem. We consider here a time-reversible
dynamical approach to Gibbs' ``microcanonical'' ( isoenergetic ) distribution for
simple systems.  To enable isoenergetic ergodicity we add occasional random rotations to the
velocities. This idea conserves energy exactly and can be made to cover the entire
energy shell with an ergodic dynamics. We entirely avoid the Poincar\'e-section holes
and island chains typical of Hamiltonian chaos. We illustrate this idea for the simplest
possible two-dimensional example, a single particle moving in a periodic square-lattice
array of scatterers, the ``cell model''.

\end{abstract}
\maketitle

\section{Introduction}

In 1984 Shuichi Nos\'e discovered a dynamics consistent with Gibbs' canonical ensemble\cite{b1,b2}
at a fixed temperature $T$. Nos\'e's idea, applied to $N$ unit-mass particles with coordinates and
momenta $\{ \ q;p \ \}$ is most simply implemented with the Nos\'e-Hoover motion equations\cite{b3}:
$$
\{ \ \dot q = p \ ; \ \dot p = F - \zeta p \ \} \ ; \ \dot \zeta = \sum^N [ \ (p^2/T) - 1 \ ]/N\tau^2 \ 
[ \ {\rm Thermostatted \ Dynamics} \ ] \ .
$$
The additional single ``thermostat variable'' $\zeta$ is a ``friction coefficient''.  When negative
it injects kinetic energy into the $(q,p)$ system.  When positive, kinetic energy is extracted.  This
integral-feedback form of the motion equations has been selected to be exactly consistent with Gibbs'
canonical distribution, $f(q,p) \propto e^{-{\cal H}/kT}$, as one of us pointed out in
1985\cite{b3}. The independent variable in the canonical ensemble is the mean value of the kinetic
energy, $K(p) = \sum (p^2/2m) \equiv \sum (p^2/2)$.   In $D$ spatial dimensions the kinetic energy
corresponds to Gibbs' ( kinetic ) temperature, $2K = DkT \equiv DT$. Throughout our work we set the
particle mass $m$ and Boltzmann's constant $k$ both equal to unity, for simplicity.

Nos\'e's original work used Hamiltonian mechanics.  It soon became evident that his thermostatted
motion equations were not ``ergodic''.  That is, the $(q,p)$ phase-space states generated by his
equations of motion failed to sample the entire phase space and instead sampled only a subspace
determined by the initial conditions.  Just as in conventional Hamiltonian mechanics two kinds of
solutions of Nos\'e's equations of motion were found, {\it regular} solutions, corresponding to
simple tori in the phase space, and {\it chaotic} Lyapunov-unstable solutions, forming a fat-fractal
``chaotic sea''.  Several years later two-thermostat motion equations were developed.  Applied to
the harmonic oscillator\cite{b4,b5} their solutions matched Gibbs' canonical distribution with an
ergodic ``chaotic sea'' and without any regular toroidal solutions.

Much later\cite{b6,b7} ergodic solutions with only a single thermostat variable were discovered. One
was found as the result of a two-parameter computerized search\cite{b6} :
$$
\{ \ \dot q = p \ ; \ \dot p = -q - 0.05\zeta p - 0.32\zeta (p^3/T) \ \} \ ;
$$
$$
\dot \zeta = 0.05[ \ (p^2/T) - 1 \ ]  + 0.32[ \ (p^4/T^2) - 3(p^2/T) \ ]
 \ [ \ {\rm 0532 \ Model} \ ] \ .
$$
A more general one-parameter thermostat, able to generate ergodic solutions to the quartic-well and
Mexican-Hat double-well problems as well as the oscillator, was developed by Tapias, Bravetti, and
Sanders as their solution of the 2016 Snook Prize problem\cite{b7} :
$$
\ddot q = -q - 2\alpha \dot q \tanh(\alpha \zeta) \ ; \ \dot \zeta = \dot q^2 - 1 \ .
$$

There are several tests that a set of motion equations must pass to establish its ergodicity.  Any
ergodic canonical dynamics algorithm must follow Gibbs' measure in phase space, $f_{Gibbs}(q,p) = 
e^{-{\cal H}/T}/\int dq\int dpe^{-{\cal H}/T}$.  Such an algorithm must also generate the canonical
averages of the various moments of the potential and kinetic energies.  For the {\it canonical}
oscillator problem the first few nonvanishing moments are the following :
$$
\langle \ q^2,p^2 \ ; \ q^4,q^2p^2,p^4 \ ; \ q^6 \dots \ \rangle =
T,T \ ; \ 3T^2,T^2,3T^2 \ ; \ 15T^3 \dots \ .
$$

\begin{figure}
\includegraphics[width=2.0in,angle=+90.,bb = 140 57 471 731]{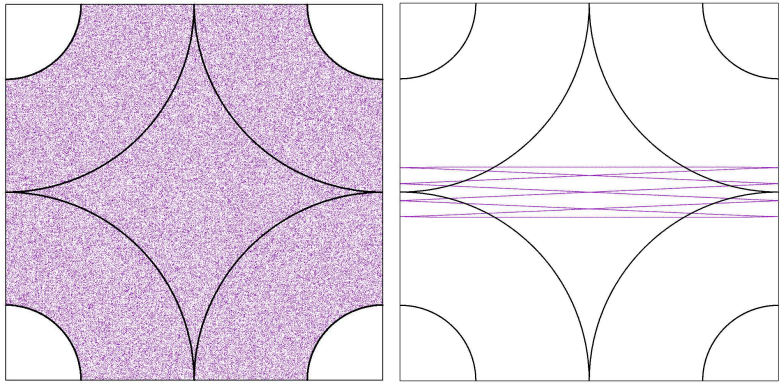}
\caption{Cell-Model dynamics with four fixed soft-disk scatterers at the corners of a $2 \times 2$
periodic cell.  For the initial condition $q = (0,0) \ ; \ p = (0.6,0.8)$ 200,000 projections from
200,000,000 timesteps, fourth-order Runge-Kutta, equally spaced in time with $\Delta t = 1000dt = 1$,
are shown at the left. At the right the initial momenta are $(0.999,\sqrt{1 - 0.999^2})$ . Evidently
this Newtonian problem is {\it not} ergodic. The black quadrant lines border the accessible regions
where the forces are nonzero. In the central diamond-shaped region the scatterer forces all vanish.
}
\end{figure}

The Kolmogorov-Arnold-Moser Theorem indicates that molecular dynamics is seldom ergodic, though the
consequences of that lack are thought to be small in most cases.  A clear lack of ergodicity is shown
in {\bf Figure 1}, where a single mass point moves in a periodic $2 \times 2$ square. Whenever the
moving particle comes closer to one of the scatterers, with $r<1$ that scatterer exerts a smooth
repulsive force :
$$
\phi(r<1) = (1-r^2)^4 \longleftrightarrow F_x = -8x(1-r^2)^3 \ ; \ F_y = -8y(1-r^2)^3 \ [ \ r < 1 \ ] \ .
$$
Although the 200,000 $(x,y)$ trajectory points appear to cover the space homogeneously at the left,
a glancing initial condition is {\it not} ergodic and generates the 200,000-point torus shown at the right.
At the right $p_x$ remains positive forever !

Ergodicity also implies that {\it any} initial condition must lead to the same time-averaged value
of the largest Lyapunov exponent, $\lambda_1 = \langle \ \lambda_1(t) \ \rangle$ .  This time-averaged
exponent describes the rate at which the separation of two nearby phase-space trajectories
tends to increase\cite{b8,b9}:
$$
\dot \delta = \lambda_1(t)\delta \ ; \ \delta \equiv \sqrt{\sum(\delta_q^2 + \delta_p^2)} \ .
$$  

\section{Ergodicity {\it via} Isoenergetic Rotations of the Momenta}

Although one might think, as we did, that {\it continuous} ( but small ) Coriolis accelerations,
$$
(\dot p_x,\dot p_y) \propto (+p_y,-p_x)
$$
would lead to ergodicity, trials of this idea were unsuccessful.  This failure led us to a successful
and simpler conservative approach to ergodicity.  We used {\it discontinuous}, rather than smooth, random
rotations of the moving particle's momentum. Let us indicate the change of momentum in two spatial
dimensions :
$$
(p_x,p_y) \propto  (\cos[\theta + \delta\theta],\sin[\theta + \delta\theta]) \ ; \
\delta\theta \propto [ \ {\cal R}-(1/2) \ ] \ {\rm with} \ {\cal R} = {\tt rund(intx,inty)} \ .
$$
The random numbers $\{ \ 0 \leq {\cal R} < 1\ \}$ come from a simple time-reversible\cite{b10}
generator :
\begin{verbatim}
      i = 1029*intx + 1731
      j = i + 1029*inty + 507*intx - 1731
      intx = mod(i,2048)
      j = j + (i - intx)/2048
      inty = mod(j,2048)
      rund = (intx + 2048*inty)/4194304.0d0
\end{verbatim}
Because only the magnitude and not the orientation of a particle's momentum contributes to the
energy such an algorithm is easily implemented.  We have confirmed that even the smallest of systems
can be made ergodic in this way, as indicated by the cell-model results in the following Section.
Because the orientations of the single-particle momenta can be chosen or changed randomly we refer to
the corresponding algorithm as a ``Monte-Carlo'' method, following the tradition of Metropolis, in
his work with the Rosenbluths and Tellers\cite{b11}.

\section{Generation of Microcanonical Poincar\'e Sections}

{\bf Figure 2} shows two $(p_x = 0)$ sections obtained from Monte-Carlo evaluations of the microcanonical
phase-space density for a moving particle with $E = \Phi + K = (1/2)$.  The cell-model potential
energy $\Phi$ describes the interaction of the moving or ``wanderer'' particle with four soft-disk
scatterers at $\{ \ \pm 1,\pm 1 \ \}$, the four vertices of a $2 \times 2$ square.  The pair potential
governing the scattering is $\phi = (1-r^2)^4$ . Rather than show numerically that the entire
microcanonical distribution is achieved in this way we will rely instead on the Metropolis,
Rosenbluths, and Tellers proof of convergence of their canonical algorithm\cite{b11} :

\begin{quote}
``Since a particle is allowed to move to any point within a square of side 2 with a finite
probability, it is clear that a large enough number of moves will enable it to reach any
point in the complete square. Since this is true of all particles, we may reach any point
in configuration space. Hence, the method is ergodic.''
\end{quote}

The simplicity of this algorithm recommends its use.

\begin{figure}
\includegraphics[width=1.5in,angle=+90.,bb= 152 85 461 712]{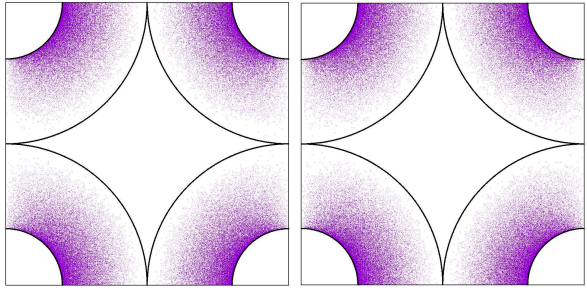}
\caption{The two $2 \times 2$ Poincar\'e sections shown here are equally likely Cell-Model states
with $p_x=0$. Unlike the uniform density of {\bf Figure 1} there is no density at all in the
central diamond, where the forces vanish, and the probability on the section is maximized along
with the magnitude of the force at the scatterer boundaries where the velocity vanishes.  Here
the initial condition corresponds to the right panel of {\bf Figure 1},
$(p_x,p_y)=(0.999,\sqrt{1 - 0.999^2})$.  The trajectory undergoes a random change of direction
every ten timesteps at the left and every thousand at the right. The Sections shown, 182,749
points at the left and 183,342 points at the right correspond to one billion timesteps,
$0 < {\rm time} < 1,000,000=1,000,000,000dt$ .
}
\end{figure}

\section{Poincar\'e Sections With and Without Rotations}
{\bf Figure 3} shows the nonergodic distribution of Poincar\'e-section points resulting from a
Runge-Kutta solution of the Hamiltonian motion equations with $dt = 0.001$ (at the left) and an
ergodic distribution resulting from occasional random rotations of the velocity vector.  The main
difference in the sections are the four missing grooves of points with $x \simeq \pm 0.17$ .  The
corresponding $p_y=0$ sections show four symmetric missing grooves with $y \simeq \pm 0.17$ .

\begin{figure}
\includegraphics[width=3.0in,angle=-90.,bb= 115 9 495 779]{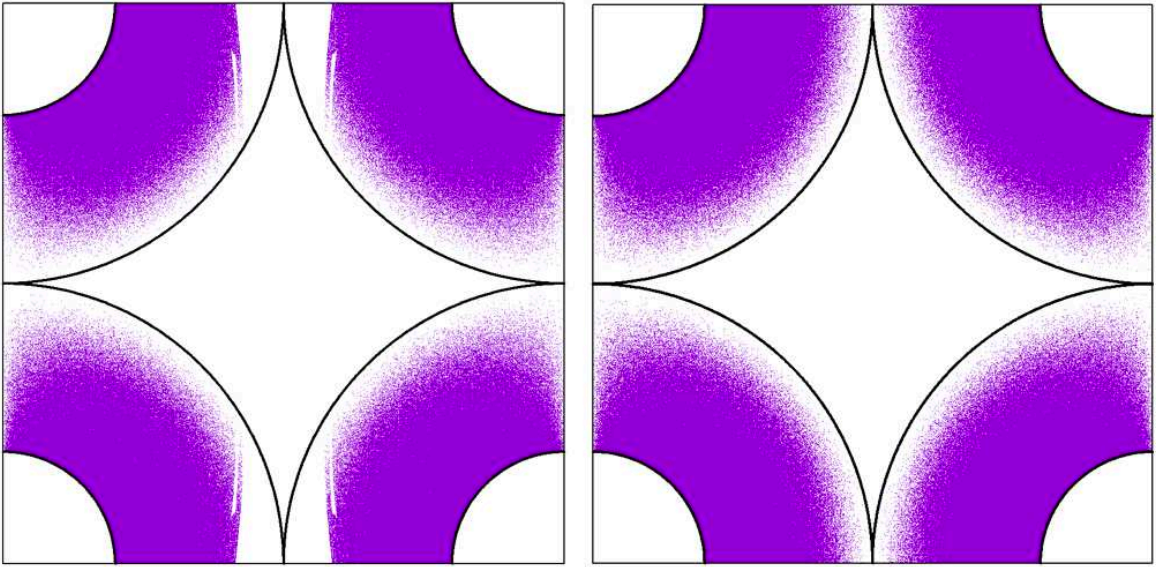}
\caption{The two $2 \times 2$ Poincar\'e sections $(x,y,p_x\equiv 0,p_y)$ shown here result from
simulations using twenty billion timesteps ( an elapsed time of $2\times 10^7$ ). At the left is
the $p_x = 0$ section according to classical mechanics with initial condition
$(x,y,p_x,p_y) = (0,0,0.6,0.8)$. At the right the velocity vector undergoes a random rotation of
${\cal R}-(1/2)$ radians after every thousand timesteps.  Notice that the grooves of excluded
states near $x \simeq \pm 0.17$ are eliminated by these rotations. Both of these sections contain
about 3.7 million points.
}
\end{figure}

\section{Conclusion}
Classical Hamiltonian mechanics is typically not ergodic.  In systems small enough for a thorough
evaluation of phase-space density simple random velocity rotations can access the missing states
which are to be expected as a consequence of the Kolmogorov-Arnold-Moser Theorem. A consequence
of this computational ergodicity is that Gibbs' statistical mechanics ( averaging over {\it all}
energy states ) can agree precisely with rotationally randomized classical mechanics.

\pagebreak


\begin{thebibliography}{99}

\bibitem{b1} S. Nos\'e, ``A Molecular Dynamics Method for Simulation in the Canonical Ensemble'',
             Molecular Physics, {\bf 52}, 255-268 (1984) \ .

\bibitem{b2} S. Nos\'e, ``A Unified Formulation of the Constant-Temperature Molecular Dynamics
             Methods'', Journal of Chemical Physics, {\bf 81}, 511-519 (1984).

\bibitem{b3} Wm. G. Hoover, ``Canonical Dynamics: Equilibrium Phase-Space Distributions'',
             Physical Review A, {\bf 31}, 1695-1697 (1985).

\bibitem{b4} Wm. G. Hoover and B. L. Holian, ``Kinetic Moments Method for the Canonical Ensemble
             Distribution'', Physics Letters A {\bf 211}, 253-257 (1996).

\bibitem{b5} G. J. Martyna, M. L. Klein, and M. Tuckerman, ``Nos\'e-Hoover Chains: the Canonical
             Ensemble {\it via} Continuous Dynamics'', The Journal of Chemical Physics {\bf 97},
             2635-2643 (1992).

\bibitem{b6} Wm. G. Hoover, C. G. Hoover, and J. C. Sprott, ``Nonequilibrium Systems: Hard Disks
             and Harmonic Oscillators Near and Far from Equilibrium'', Molecular Simulation,
             {\bf 42}, 1300-1316 (2016).

\bibitem{b7} D. Tapias, A. Bravetti, and D. P. Sanders, ``Ergodicity of One-Dimensional Systems
             Coupled to the Logistic Thermostat'', Computational Methods in Science and Technology
             {\bf 23}, 11-18 (2017) = arXiv 1611.05090 .

\bibitem{b8}  I. Shimada and T. Nagashima, ``A Numerical Approach to Ergodic Problems of Dissipative
              Dynamical Systems'', Progress of Theoretical Physics {\bf 61}, 1605-1616 (1979).

\bibitem{b9}  G. Benettin, L. Galgani, A. Giorgilli, and J.-M. Strelcyn, ``Lyapunov Characteristic 
              Exponents for Smooth Dynamics Systems and for Hamiltonian Systems; a Method for
              Computing All of Them, Parts I and II: Theory and Numerical Application'', Meccanica
              {\bf 15}, 9-20 and 21-30 (1980).

\bibitem{b10} F. Ricci-Tersenghi, ``The Solution to the Challenge in `Time-Reversible Random Number
              Generators' '' ar$\chi$iv:1305.1805 (2013).

\bibitem{b11} N. Metropolis, A. W. and M. N. Rosenbluth, and A. H. and E. Teller, ``Equation of State
              Calculations by Fast Computing Machines'', The Journal of Chemical Physics {\bf 21},
              1087-1092 (1953).


\end{thebibliography}
\end{document}